\documentclass[letter,twocolumn]{jpsj3}

\usepackage{txfonts}

\usepackage{amsmath}
\usepackage{amsfonts}
\usepackage{bm}
\usepackage{graphicx}
\usepackage{color}

\title{Path-integral Monte Carlo study on a droplet of a dipolar
Bose-Einstein condensate stabilized by quantum fluctuation
}

\author{Hiroki Saito}
\inst{Department of Engineering Science, University of
Electro-Communications, Tokyo 182-8585, Japan}

\abst{Motivated by the recent experiments [H. Kadau {\it et al.},
Nature (London) {\bf 530}, 194 (2016); I. Ferrier-Barbut {\it et al.},
arXiv:1601.03318] and theoretical prediction (F. W\"achtler and L. Santos,
arXiv:1601.04501), the ground state of a dysprosium Bose-Einstein
condensate with strong dipole-dipole interaction is studied using the
path-integral Monte Carlo method.
It is shown that quantum fluctuation can stabilize the condensate against
dipolar collapse.
}

\begin{document}
\maketitle

Realization of Bose-Einstein condensates (BECs) of atoms with large
dipole-dipole interaction (DDI)~\cite{Gries,Lu,Aikawa} has opened up the
physics of ferromagnetic superfluidity.
Experimental researches have been focused on the long-range and
anisotropic nature of the DDI, such as anisotropic
deformation~\cite{Stuhler,Lahaye,Koch} and
excitation~\cite{Bismut,Bismut2} of the cloud, anisotropic collapse and
expansion~\cite{Lahaye2,Metz}, and spinor-dipolar
effects~\cite{Pasquiou,Pasquiou2,Eto}.

Recently, the experimental group in Stuttgart observed~\cite{Kadau,Barbut}
droplet lattice formation in a BEC of $^{164}{\rm Dy}$ atoms, which have
magnetic moment much larger than alkali atoms.
A pancake-shaped BEC of $^{164}{\rm Dy}$ atoms is prepared for a
scattering length larger than the critical value for the dipolar
collapse.
The scattering length is then decreased to below the critical value for
the collapse using Feshbach resonance, and the system becomes unstable due
to the attractive part of the DDI.
An instability, similar to the Rosensweig instability~\cite{Cowley,Saito}
in magnetic liquids, splits the condensate into droplets, and they form a
stable triangular lattice.

Theoretical studies have been performed to explain the observation in the
Stuttgart experiment.
However, it has been found that simple mean-field theory, i.e., the
Gross-Pitaevskii (GP) equation with DDI, cannot reproduce the experimental
results;
numerical studies of the GP equation have shown that the droplets always
collapse immediately after they form, since the quantum pressure and
$s$-wave repulsive interaction cannot support the attractive force of 
the DDI.
To solve this problem, is was proposed that the droplets can be stabilized
if large three-body repulsion exists.
It was shown that the GP equation with appropriate strength of three-body
repulsion can reproduce the experimental results~\cite{KuiTian,Bisset}.
Very recently, another mechanism to explain the stable droplets was
proposed:
W\"achtler and Santos~\cite{Wachtler} showed that the GP equation with
a Lee-Huang-Yang (LHY)~\cite{Lee} correction term can stabilize the
droplets and reproduce the experimental results.

Motivated by the theoretical prediction in Ref.~\citen{Wachtler}, in this
Letter, we examine whether the quantum fluctuation can stabilize the
droplet against dipolar collapse, using the path-integral Monte Carlo
(PIMC) approach~\cite{Ceperley}.
Many researchers have employed the PIMC method to explore the quantum
many-body properties of ultracold
atoms~\cite{Krauth,Gruter,Holzmann,Holzmann2,Nho,Nho_dipole,Pilati,Holzmann08,Filinov}.
We will show that a stable droplet state is obtained by the PIMC method,
even when the ground state does not exist and the collapse occurs in the
simple mean-field theory.
The density profiles of the atomic clouds obtained by the PIMC method are
compared with those by the GP equation with LHY correction proposed in
Ref.~\citen{Wachtler}.

We consider a system of $^{164}{\rm Dy}$ atoms with mass $m$ confined in a
trap potential $V(\bm{r})$, in which the direction of the magnetic dipole
moment of the atoms is fixed to the $z$ axis.
The Hamiltonian for the system is given by
\begin{equation} \label{H}
H = \sum_{j=1}^N \left[ \frac{\bm{p}_j^2}{2m} + V(\bm{r}_j) \right]
+ \sum_{j_1 < j_2} U(\bm{r}_{j_1} - \bm{r}_{j_2}),
\end{equation}
where $N$ is the number of atoms, and $\bm{r}_j$ and $\bm{p}_j$ are the
position and momentum operators of the $j$th atom.
The system is confined in a harmonic potential $V(\bm{r}) = m
(\omega_x^2 x^2 + \omega_y^2 y^2 + \omega_z^2 z^2) / 2$, where $\omega_x$,
$\omega_y$, and $\omega_z$ are the trap frequencies.
The interaction $U$ between atoms consists of the hard-sphere potential
with a radius $a$ and the magnetic DDI as
\begin{equation}
U(\bm{r}) = U_{\rm hard}(r) + \frac{\mu_0 \mu^2}{4\pi} \frac{1 -
3\cos\chi}{r^3},
\end{equation}
where $U_{\rm hard}(r) = \infty$ for $r < a$ and $U_{\rm hard}(r) = 0$ for
$r > a$, $\mu_0$ is the magnetic permeability of the vacuum, $\mu = 9.93
\mu_B$ is the magnetic dipole moment of a $^{164}{\rm Dy}$ atom with
$\mu_B$ being the Bohr magneton, and $\chi$ is the angle between $\bm{r}$
and the $z$ axis.
It is known that the $s$-wave scattering length coincides with the radius
$a$ of the hard-sphere potential.

In thermal equilibrium at temperature $T$, the probability that the atoms
are located at $R = \{\bm{r}_1, \bm{r}_2, \cdots, \bm{r}_N \}$ is
proportional to $\sum_P \langle R | e^{-\beta H} | P R \rangle$, where $P$
represents permutation of indices to assure the Bose symmetry and $\beta =
1 / (k_B T)$ with $k_B$ being the Boltzmann constant.
The bracket $\langle R | e^{-\beta H} | P R \rangle$ is divided into the
path-integral form,
\begin{eqnarray} \label{path}
& & \int \cdots \int dR_1 dR_2 \cdots dR_{M-1}
\langle R | e^{-\beta H / M} | R_1 \rangle
\nonumber \\
& &  \times \langle R_1 | e^{-\beta H / M} | R_2 \rangle \cdots
\langle R_{M-1} | e^{-\beta H / M} | PR \rangle,
\end{eqnarray}
where $M$ is the number of ``slices''.
Each bracket in Eq.~(\ref{path}) is approximated by
\begin{equation} \label{prop}
\langle R | e^{-\beta H / M} | R' \rangle \simeq
P_{\rm noint} P_{\rm hard} P_{\rm ddi}.
\end{equation}
The part of noninteracting particles in a harmonic potential has the
form~\cite{Feynman},
\begin{eqnarray}
& & P_{\rm noint}(R, R'; \tau)
= \prod_{j=1}^N \prod_{\sigma=x,y,z} \left( \frac{m
\omega_\sigma}{2\pi\hbar \sinh \omega_\sigma \tau} \right)^{1/2}
\nonumber \\
& & \times \exp\left\{ -\frac{m \omega_\sigma}{2 \hbar \sinh \omega_\sigma
\tau} \left[ (\sigma_j^2+\sigma'^{2}_j) \cosh \omega_\sigma \tau - 2
\sigma_j \sigma'_j \right] \right\},
\nonumber \\
\end{eqnarray}
where $\tau = \hbar \beta / M$.
For the hard-sphere interaction part in Eq.~(\ref{prop}), we adopt the
expression derived in Ref.~\citen{Cao},
\begin{eqnarray}
& & P_{\rm hard}(R, R'; \tau) \nonumber \\
& & = \prod_{j_1 > j_2} \Biggl\{ 1 - \frac{a
(\rho_{12} + \rho'_{12} - a)}{\rho_{12} \rho'_{12}}
\nonumber \\
& & \times \exp\left[ -\frac{1}{2\tau} (\rho_{12} - a)(\rho'_{12} - a)
\left( 1 + \frac{\bm{\rho}_{12} \cdot \bm{\rho}_{12}'}{\rho_{12}
\rho'_{12}} \right) \right] \Biggr\}
\nonumber \\
& & \times H(\rho_{12} - a) H(\rho'_{12} - a),
\end{eqnarray}
where $\bm{\rho}_{12} = \bm{r}_{j_1} - \bm{r}_{j_2}$, 
$\bm{\rho}'_{12} = \bm{r}'_{j_1} - \bm{r}'_{j_2}$,
and $H$ is the Heaviside step function, i.e., $P_{\rm hard}$ vanishes
when the distance between any two particles is less than $a$.
The DDI part in Eq.~(\ref{prop}) is approximated as
\begin{eqnarray} \label{Pddi}
& & P_{\rm ddi}(R, R'; \tau)
\nonumber \\
& & = \exp\left\{ -\frac{\tau}{\hbar} \frac{\mu_0 \mu^2}{4\pi}
\frac{1}{2} \sum_{j_1 <j_2}
\left[ \frac{1 - 3\cos\theta_{12}}{f_{\rm cutoff}(\rho_{12})} + \frac{1 -
3\cos\theta'_{12}}{f_{\rm cutoff}(\rho'_{12})} \right] \right\},
\nonumber \\
\end{eqnarray}
where $f_{\rm cutoff}(r) = r^3$ for $r > R_{\rm cutoff}$ and
$f_{\rm cutoff}(r) = R_{\rm cutoff}^3$ for $r < R_{\rm cutoff}$. 
The cutoff radius $R_{\rm cutoff}$ is introduced to avoid the steep
increase in the DDI potential near $r = 0$, which ruins the calculation.
The validity of the cutoff will be discussed later.
Using these expressions of $P_{\rm noint}$, $P_{\rm hard}$,
and $P_{\rm ddi}$, $\sum_P \langle R | e^{-\beta H} | P R \rangle$ is
evaluated by the multilevel Metropolis sampling~\cite{Ceperley}, where
$R_1, R_2, \cdots, R_{M-1}$, and $R$ are sampled with an appropriate
probability.
Taking the average of $R$, one obtains the density distribution
$n(\bm{r})$ of the atomic cloud in thermal equilibrium.
Typically, after $10^3$-$10^4$ Monte Carlo sweeps are performed to relax
the system, $10^3$-$10^4$ samples are taken for the average.

Before showing the PIMC results, we briefly review the mean-field theory
with the LHY correction proposed in Ref.~\citen{Wachtler}.
The LHY correction of the chemical potential in a homogeneous dipolar BEC
with density $n$ is given
by~\cite{Lima}
\begin{equation} \label{mu}
\Delta \mu(n) = \frac{32}{3\sqrt{\pi}} g n \sqrt{n a^3} F(\epsilon_{dd}),
\end{equation}
where $g = 4\pi\hbar^2 a / m$, $\epsilon_{dd} = \mu_0 \mu^2 / (3g)$, and
\begin{equation} \label{F}
F(\epsilon_{dd}) = \frac{1}{2} \int_0^\pi d\theta \sin\theta [1 +
\epsilon_{dd} (3 \cos^2\theta - 1)]^{5/2}.
\end{equation}
The integral in Eq.~(\ref{F}) is taken for the range in which the
integrand is real.
Using the local density approximation, the LHY correction in
Eq.~(\ref{mu}) is incorporated into the GP equation, giving
\begin{eqnarray} \label{GPLHY}
i \hbar \frac{\partial \psi}{\partial t} & = & \Biggl[ -\frac{\hbar^2}{2m}
\nabla^2 + V + g |\psi|^2 \psi
\nonumber \\
& & + \frac{\mu_0\mu^2}{4\pi} \int d\bm{r'}
\frac{1 - 3 \cos\chi'}{|\bm{r} - \bm{r}'|^3} |\psi(\bm{r}')|^2
+ \Delta\mu(|\psi|^2) \Biggr] \psi,
\nonumber \\
\end{eqnarray}
where $\chi'$ is the angle between $\bm{r} - \bm{r}'$ and the $z$ axis.
The macroscopic wave function $\psi$ is normalized as $\int |\psi|^2
d\bm{r} = N$.
The DDI energy and the LHY correction $\Delta \mu$ are roughly
proportional to $|\psi|^2$ and $|\psi|^3$, respectively.
Therefore, when the peak density is increased by the DDI, the energy is
dominated by the LHY correction term, which stops the collapse.
The LHY quantum fluctuation can thus prevent the collapse and stabilize
droplets.
In the following results, the stationary states of the GP equation are
obtained by the imaginary-time propagation method, in which $i$ on the
left-hand side is replaced with $-1$ and the wave function is normalized
in every time step.

\begin{figure}
\begin{center}
\includegraphics[width=8.5cm]{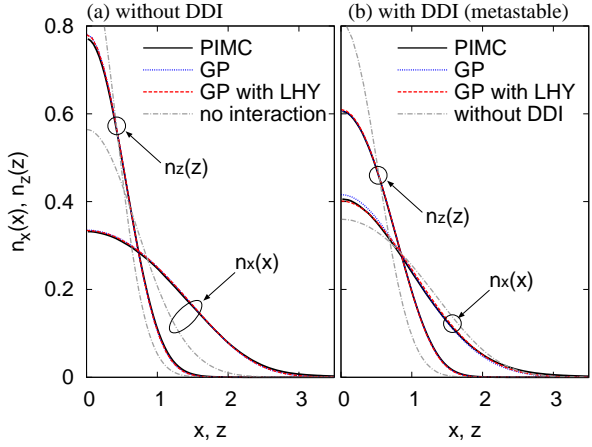}
\end{center}
\caption{
Integrated density distributions $n_x(x)$ and $n_z(z)$ obtained by the
path-integral Monte Carlo (PIMC) method (black solid curves) and the
Gross-Pitaevskii (GP) equations without (blue dotted curves) and with (red
dashed curves) the Lee-Huang-Yang (LHY) correction in Eq.~(\ref{GPLHY}).
$N = 1024$ atoms are confined in a harmonic potential with
frequencies $(\omega_x, \omega_y, \omega_z) = 2\pi \times (46, 44, 133)$.
(a) Density distributions without the dipole-dipole interaction (DDI),
where $a = 100 a_0$ and $M = 256$.
The gray dot-dashed curves show the harmonic oscillator ground state.
(b) Density distributions of the metastable state with DDI, where $a = 70
a_0$, $M = 256$, and $R_{\rm cutoff} = 0.2 a_x$.
For this value of $a$, the GP equation has a metastable state.
The gray dot-dashed curves are obtained by the GP equation without DDI and
LHY correction.
The units of length and density distribution are $a_x = [\hbar / (m
\omega_x)]^{1/2}$ and $a_x^{-1}$.
}
\label{f:stable}
\end{figure}
We first check that the PIMC method reproduces the mean-field theory, when
the LHY correction is small.
We consider a system of $N = 1024$ atoms confined in a harmonic trap with
frequencies $(\omega_x, \omega_y, \omega_z) = 2\pi \times (46, 44, 133)$
Hz~\cite{Kadau}.
We take $\hbar \bar\omega \beta \equiv \hbar (\omega_x \omega_y
\omega_z)^{1/3} \beta = 0.3$, which corresponds to $T \simeq 7.4$ nK.
The critical temperature for Bose-Einstein condensation of an ideal Bose
gas is $T_c \simeq 0.94 \hbar \bar\omega N^{1/3} / k_B \simeq 29$ nK.
Figure~\ref{f:stable}(a) shows the results without DDI, where $a = 100
a_0$ with $a_0$ being the Bohr radius.
We define the integrated density distributions as
\begin{equation}
n_x(x) = \frac{1}{N} \int n(\bm{r}) dy dz, \;
n_z(z) = \frac{1}{N} \int n(\bm{r}) dx dy,
\end{equation}
where $n(\bm{r})$ is the atom density.
In Fig.~\ref{f:stable}(a), the density profiles obtained by the PIMC
method almost agree with those by the GP equation.
For these parameters, the GP results with and without the LHY correction
cannot be discerned.

Figure~\ref{f:stable}(b) shows the result with DDI for $a = 70 a_0$.
For this value of $a$, the relative strength of the DDI is $\epsilon_{dd}
\simeq 1.87$ and the GP equation without the LHY correction has a
metastable state, where the energy barrier originates from the quantum
pressure~\cite{Koch}.
Due to the anisotropic nature of the DDI, the atomic cloud is slightly
elongated in the $z$ direction and shrunk in the $x$-$y$ direction,
compared with that without DDI.
We see that the density distribution obtained by the PIMC method is in
good agreement with those by the GP equation with DDI.
This indicates that the PIMC method can be used to obtain not only the
ground state but also a metastable state.
The cutoff radius used in Fig.~\ref{f:stable}(b) is $R_{\rm cutoff} =
0.2 a_x$, where $a_x = [\hbar / (m \omega_x)]^{1/2}$.
Almost the same result is obtained for $R_{\rm cutoff} = 0.1 a_x$.

We next examine whether the quantum fluctuation can stop the dipolar
collapse.
The state in Fig.~\ref{f:stable}(b) is the metastable state, and beyond
the energy barrier, the energy of the atomic cloud decreases as it
shrinks.
If the LHY correction is absent, the GP equation has no lower energy bound
and the peak density diverges; there is no ground state.
The LHY correction in Eq.~(\ref{GPLHY}) suppresses the divergence of the
peak density and allows the ground state~\cite{Wachtler}.
To cross the energy barrier in the numerical calculations, the radial
harmonic frequencies $\omega_x$ and $\omega_y$ are temporarily increased
during Monte Carlo sweeps in the PIMC and during imaginary-time
propagation in the GP equation, which shrinks the atomic cloud in the
$x$-$y$ directions.
Starting from these states, the system goes to the ground state beyond the
energy barrier.

\begin{figure}
\begin{center}
\includegraphics[width=8.5cm]{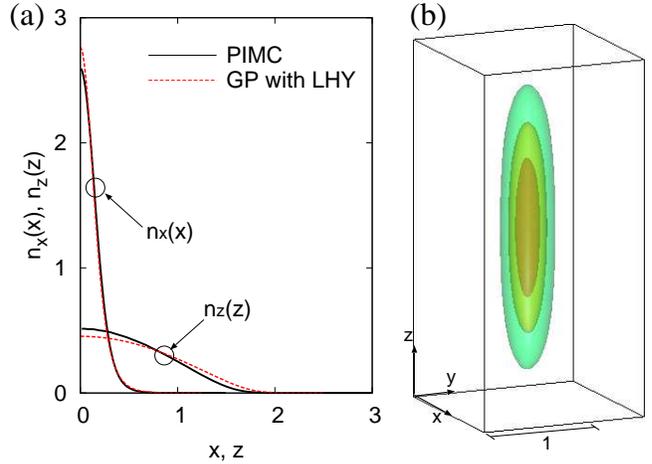}
\end{center}
\caption{
(a) Integrated density distributions $n_x(x)$ and $n_z(z)$ obtained by the
PIMC method (black solid curves) and the GP equation with LHY correction
in Eq.~(\ref{GPLHY}) (red dashed curves).
The parameters are the same as those in Fig.~\ref{f:stable}(b).
There is no stable ground state in the GP equation without LHY correction
for these parameters.
(b) Isodensity surfaces obtained by the PIMC method.
The surfaces represent $1/5$, $2/5$, and $3/5$ of the peak density $\simeq
3 \times 10^{15}$ ${\rm cm}^{-3}$.
The size of the frame is $1.5 \times 1.5 \times 3$.
The units of length and density distribution are $a_x = [\hbar / (m
\omega_x)]^{1/2}$ and $a_x^{-1}$.
}
\label{f:stabilized}
\end{figure}
Figure~\ref{f:stabilized} shows the density distributions of the state
that has crossed the energy barrier, where the parameters are the same as
those in Fig.~\ref{f:stable}(b).
From Fig.~\ref{f:stabilized}(a), we find that both PIMC method and GP
equation with LHY correction provide stable states, in which the dipolar
collapse is suppressed and the peak density is kept finite.
The density distribution obtained by the PIMC method slightly deviates
from that by the GP equation with LHY correction, mainly due to the errors
in the PIMC, which will be explained later.
The GP equation with LHY correction may also be inaccurate due to the
local density approximation.
Figure~\ref{f:stabilized}(b) shows the isodensity surfaces of the
three-dimensional density distribution obtained by the PIMC method.
The atomic cloud is highly deformed to the cigar shape by the anisotropic
DDI, while the trap potential is pancake shaped.
The peak density in Fig.~\ref{f:stabilized} is $\sim 3 \times 10^{15}$
${\rm cm}^{-3}$ and the gas parameter is $n a^3 \sim 10^{-4}$.
The three-body recombination is expected to occur predominantly at the
density peak, which is the reason for the atomic loss observed in the
experiment~\cite{Kadau}.

\begin{figure}
\begin{center}
\includegraphics[width=8.5cm]{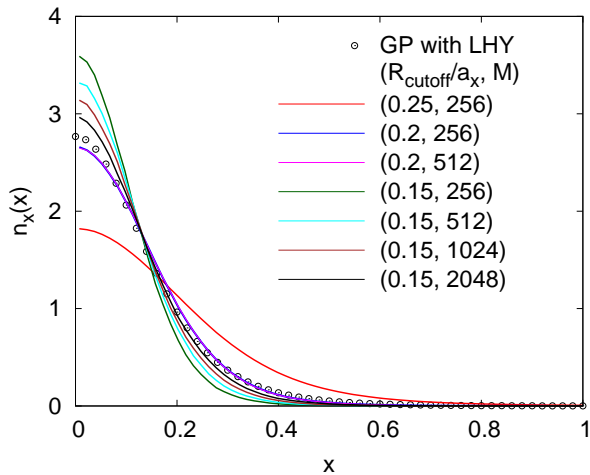}
\end{center}
\caption{
Dependence of the density distribution $n_x(x)$ on the cutoff radius
$R_{\rm cutoff}$ and the number of slices $M$ in the PIMC method.
The value of $a = 70 a_0$ is the same as that in Fig.~\ref{f:stabilized}.
From top to bottom of the peak values $n_x(x = 0)$, $(R_{\rm cutoff} /
a_x, M) = (0.15, 256)$, $(0.15, 512)$, $(0.15, 1024)$, $(0.15, 2048)$,
$(0.2, 256)$, $(0.2, 512)$, and $(0.25, 256)$. 
The circles are obtained by the GP equation with LHY correction.
The units of length and density distribution are $a_x = [\hbar / (m
\omega_x)]^{1/2}$ and $a_x^{-1}$.
}
\label{f:rmdep}
\end{figure}
We check the validity of the cutoff made in the DDI potential in
Eq.~(\ref{Pddi}).
Figure~\ref{f:rmdep} shows the dependence of the density distribution
$n_x(x)$ on the cutoff radius $R_{\rm cutoff}$ and the number of slices
$M$ in the PIMC method.
For $R_{\rm cutoff} = 0.25 a_x$, the distribution $n_x(x)$ is
substantially wider than others, and we presume that $R_{\rm cutoff} =
0.25 a_x$ is too large to give the accurate result. 
For $R_{\rm cutoff} = 0.2 a_x$, $n_x(x)$ of the PIMC is close to that of
the GP equation with LHY correction.
Since the results of $M = 256$ and $M = 512$ are almost the same, the
number of slices is enough.
However, for $R_{\rm cutoff} = 0.15 a_x$, $n_x(x)$ significantly depends
on $M$, when $M$ is inadequate;
$M$ must be 2048 or larger.
Therefore, the number of slices $M$ must be increased with a decrease in
$R_{\rm cutoff}$.
The computational amount is proportional to $M$, and the calculation for
$R_{\rm cutoff} < 0.15 a_x$ is extremely difficult.
It seems that the density distribution converges to around that of the GP
equation with LHY correction.

The accuracy of the present PIMC calculation is thus restricted by the
the primitive approximation of $P_{\rm ddi}$ in Eq.~(\ref{Pddi}),
whose $r^{-3}$ steepness requires large numbers of slices $M$ and Monte
Carlo samplings.
A more suitable expression for $P_{\rm ddi}$ is needed.
Another bottleneck is the long-range nature of the DDI, which costs
$O(N^2)$ calculations per Monte Carlo sweep.
The $O(N)$ Monte Carlo technique~\cite{Fukui} may circumvent this
problem.
With these improvements, it may be possible not only to perform more
accurate calculation, but also to simulate the droplet pattern formation
observed in the experiment with $N \sim 10^5$ atoms~\cite{Kadau}.
The results obtained by the PIMC method should be compared with other
methods, such as the diffusion Monte Carlo method.

In conclusion, we have investigated the stability of a strong dipolar BEC
against collapse, motivated by the recent experiments~\cite{Kadau,Barbut}
and the theoretical prediction~\cite{Wachtler}.
Using the PIMC method, we showed that the system has a stable ground state
even in the parameters for which the simple GP equation cannot sustain the
system against dipolar collapse, which implies that the quantum
fluctuation stabilizes the system.
We compared the PIMC results with those obtained by the GP equation with
the LHY correction proposed in Ref.~\citen{Wachtler}, and found that they
are in qualitative agreement.
The present results indicate that the quantum fluctuation plays an
important role in the droplet stabilization observed in the
experiments~\cite{Kadau,Barbut}.

\begin{acknowledgments}
I wish to thank Kui-Tian Xi for fruitful discussion.
This work was supported by JSPS KAKENHI Grant Number 26400414 and by MEXT
KAKENHI Grant Number 25103007.
\end{acknowledgments}

\end{document}